\documentclass[12pt]{article}
\usepackage{t1enc}
\usepackage[utf8]{inputenc}
\usepackage[english]{babel}
\usepackage{graphicx}
\usepackage{amssymb}
\usepackage{amsmath}
\usepackage{amsthm}
\usepackage{braket}

\def\UN{\mathbf{1}}

\def\UN{\mathbf{1}}

\def\cros{\raise1.9pt\hbox{$\scriptscriptstyle
          >$}\!\raise1.5pt\hbox{$\scriptstyle\triangleleft\,$}}

\def\S{\Sigma}
\def\s{\sigma}
\def\cP{{\cal P}}

\def\l{{\lambda}}

\def\O{\Omega}

\def\w{\wedge}

\theoremstyle{definition}
\theoremstyle{definition}
\theoremstyle{definition}
\theoremstyle{definition}
\newcommand{\noi}{\vspace{0.1in} \noindent}

\title{\bf On the three types of Bell's inequalities}
\author{\textit{Gábor Hofer-Szabó}\thanks{Research Center for the Humanities, Budapest, email: szabo.gabor@btk.mta.hu}}
\date{}

\begin{document}
\maketitle

\begin{abstract}
Bell's inequalities can be understood in three different ways depending on whether the numbers featuring in the inequalities are interpreted as classical probabilities, classical conditional probabilities, or quantum probabilities. In the paper I will argue that the violation of Bell's inequalities has different meanings in the three cases. In the first case it rules out the interpretation of certain numbers as probabilities of events. In the second case it rules out a common causal explanation of conditional correlations of certain events (measurement outcomes) conditioned on other events (measurement settings). Finally, in the third case the violation of Bell's inequalities neither rules out the interpretation of these numbers as probabilities of events nor a common causal explanation of the correlations between these events---provided both the events and the common causes are interpreted non-classically.
\vspace{0.1in}

\end{abstract}

\section{Introduction}

Ever since its appearance three decades ago, Itamar Pitowsky's \textit{Quantum Probability -- Quantum Logic} has been serving as a towering lighthouse showing the way for many working in the foundations of quantum mechanics. In this wonderful book Pitowsky provided an elegant geometrical representation of classical and quantum probabilities: a powerful tool in tackling many difficult formal and conceptual questions in the foundations of quantum theory. One of them is the meaning of the violation of Bell's inequalities.  

However, for someone reading the standard foundations of physics literature on Bell's theorems, it is not easy to connect up
Bell's inequalities as they are presented  in Pitowsky's book with the inequalities presented by Bell, Clauser, Horne, Shimony etc. The main difference, to make it brief, is that Bell's inequalities in their original form are formulated in terms of \textit{conditional} probabilities, representing certain measurement outcomes provided that certain measurement are performed, while Pitowsky's Bell inequalities are formulated in terms of \textit{unconditional} probabilities, representing the distribution of certain underlying properties or events responsible for the measurement outcomes. 

Let us see this difference in more detail. Probabilities enter into quantum mechanics \textit{via} the trace formula
\begin{eqnarray}\label{qprob}
p &=& \mbox{Tr}(\hat{\rho}\hat{A})
\end{eqnarray}
where $\hat{\rho}$ is a density operator, $\hat{A}$ is the spectral projection associated to eigenvalue $\alpha$ of the self-adjoint operator $\hat{a}$, and Tr is the trace function. Let us call the probabilities generated by the trace formula (\ref{qprob}) \textit{quantum probabilities}.

Now, what is the physical interpretation of quantum probabilities? There are two possible answers to this question according to two different interpretations of quantum mechanics: 

\begin{enumerate}
\item On the \textit{operational (minimal) interpretation}, the density operator $\hat{\rho}$ represents the state or preparation $s$ of the system; the self-adjoint operator $\hat{a}$ represents the measurement $a$ performed on the system; and the spectral projection $\hat{A}$ represents the outcome $A$ of the measurement $a$. On this interpretation the quantum probability is interpreted as the \textit{conditional probability}
\begin{eqnarray}\label{condprob}
p &=& p_s(A|a) 
\end{eqnarray}
that is the probability of obtaining outcome $A$ \textit{provided} that the measurement $a$ has been performed on the system previously prepared in state $s$. This interpretation is called \textit{minimal} since the fulfillment of (\ref{condprob}) is a necessary condition for the theory to be empirically adequate. 

\item On the \textit{ontological (deterministic hidden variable, property) interpretation}, the density operator $\hat{\rho}$ represents the distribution $\rho$ of the ontological states $\l \in \Lambda$ in the preparation $s$; the operator $\hat{a}$ represents the physical magnitude $a^*$; and the projection $\hat{A}$ represents the event $A^*$ that the value of $a^*$ is $A^*$.\footnote{I denote the event and the value by the same symbol.} Denote by $\Lambda^{A^*} = \{\lambda^{A^*}\}$ the set of those ontological states for which the value  $a^*$ is $A^*$. On the ontological interpretation the quantum probability is (intended to be) interpreted as the \textit{unconditional probability}
\begin{eqnarray}\label{prob}
p &=& p_s(A^*) = \int_{\Lambda^{A^*}} \rho(\l) \, d\l
\end{eqnarray}
that is the probability of the event $A^*$ in the state $s$.

The ontological and the operational definition are connected as follows. The measurement $a$ is said to measure the physical magnitude $a^*$ only if the following holds: the outcome of measurement $a$ performed on the system will be $A$ if and only if the system is in an ontological state for which the value of $a^*$ is $A^*$:
\begin{eqnarray*}
p_\l(A|a) = 1 \quad \quad \quad \mbox{if and only if} \quad \quad \quad \l \in \Lambda^{A^*}
\end{eqnarray*}
\end{enumerate}
As is seen, the two interpretations differ in how they treat quantum probabilities. On the operational interpretation quantum probabilities are condition probabilities, while on the ontological interpretation they are unconditional probabilities. (Note that in both interpretations probabilities are ``conditioned'' on the preparation which will be dropped from the next section.)

Now, consider a \textit{set} of self-adjoint operators $\{\hat{a}_i\}$ ($i \in I$) each with two spectral projections $\{\hat{A}_i, \hat{A}^\perp_i\}$. Correspondingly, consider a set of measurements $\{a_i\}$ each with two outcomes $\{A_i, A^\perp_i\}$ and a set of magnitudes $\{a^*_i\}$ each with two values $\{A^*_i, A^{*\perp}_i\}$. If $\hat{A}_i$ and $\hat{A}_j$ are commuting, then the quantum probabilities
\begin{eqnarray}
p_i &=& \mbox{Tr}(\hat{\rho}\hat{A}_i) \label{qprob1} \\
p_{ij} &=& \mbox{Tr}(\hat{\rho}\hat{A}_i\hat{A}_j) \label{qprob2} 
\end{eqnarray}
can be interpreted either operationally: 
\begin{eqnarray}
p_i &=& p_s(A_i|a_i) \label{condprob1} \\
p_{ij} &=& p_s(A_i \w A_j |a_i \w a_j) \label{condprob2}
\end{eqnarray}
or ontologically:
\begin{eqnarray}
p_i &=& p_s(A^*_i) \label{prob1} \\
p_{ij} &=& p_s(A^*_i \w A^*_j) \label{prob2}
\end{eqnarray}

Consider a paradigmatic Bell inequality, the Clauser-Horne inequalities:
\begin{eqnarray}
-1 \leqslant  p_{ij}+p_{i'j}+p_{ij'}-p_{i'j'}-p_{i}-p_{j} \leqslant 0&\quad & i,i'=1,2;\,\, j,j'=3,4;\,\, i\neq i';\,\, j\neq j' \quad  \label{Clauser-Horne}
\end{eqnarray}
The numbers featuring in (\ref{Clauser-Horne}) are probabilities. But which type of probabilities? Are they simply (uninterpreted) quantum probabilities of type (\ref{qprob1})-(\ref{qprob2}), or condition probabilities of type (\ref{condprob1})-(\ref{condprob2}), or unconditional probabilities of type (\ref{prob1})-(\ref{prob2})? Depending on how the probabilities in the Clauser-Horne inequalities are understood, the inequalities can be written out in the following three different forms:
\begin{eqnarray}
-1 \leqslant  \mbox{Tr}(\hat{\rho}\hat{A}_i \hat{A}_j) + \mbox{Tr}(\hat{\rho}\hat{A}_{i'} \hat{A}_j) + \mbox{Tr}(\hat{\rho}\hat{A}_i \hat{A}_{j'}) \nonumber \\
- \mbox{Tr}(\hat{\rho}\hat{A}_{i'} \hat{A}_{j'}) - \mbox{Tr}(\hat{\rho}\hat{A}_i) - \mbox{Tr}(\hat{\rho}\hat{A}_j) \leqslant 0 \label{Clauser-Horne1}
\end{eqnarray}
(where $\hat{A}_i \hat{A}_j, \hat{A}_{i'} \hat{A}_j, \hat{A}_i \hat{A}_{j'}$ and $\hat{A}_{i'} \hat{A}_{j'}$ are commuting projections); or
\begin{eqnarray}
-1 \leqslant  p_s(A_i \w A_j|a_i \w a_j) + p_s(A_{i'} \w A_j | a_{i'} \w a_j)+ p_s(A_i \w A_{j'} | a_i \w a_{j'})  \nonumber \\
- p_s(A_{i'} \w A_{j'} | a_{i'} \w a_{j'}) - p_s(A_i | a_i) -p_s(A_j | a_j) \leqslant 0 \label{Clauser-Horne2}
\end{eqnarray}
or
\begin{eqnarray}
-1 \leqslant  p_s(A^*_i \w A^*_j) + p_s(A^*_{i'} \w A^*_j)+ p_s(A^*_i \w A^*_{j'}) \nonumber \\
- p_s(A^*_{i'} \w A^*_{j'}) - p_s(A^*_i) -p_s(A^*_j) \leqslant 0 \label{Clauser-Horne3}
\end{eqnarray}

Thus, altogether we have three different types of Bell inequalities depending on whether and how the probabilities featuring in (\ref{Clauser-Horne}) are physically interpreted. The aim of the paper is to clarify as to what is exactly excluded if Bell's inequalities of type (\ref{Clauser-Horne1}), (\ref{Clauser-Horne2}) or (\ref{Clauser-Horne3}) are violated. I will argue that the violation has three different meanings in the three different cases. In case of (\ref{Clauser-Horne3}), 
when the probabilities are classical unconditional probabilities, it rules out the interpretation of certain numbers as probabilities of events or properties. These are the inequalities which Pitowsky identified and categorized. The violation of inequalities (\ref{Clauser-Horne2}), when the probabilities are classical conditional probabilities, does not rule out the interpretation of certain numbers as conditional probabilities of events but only a common causal explanation of the conditional correlations between these events. These are the Bell inequalities used in the standard foundations of physics literature. Finally, the violation of Bell's inequalities (\ref{Clauser-Horne1}) neither rules out the interpretation of certain numbers as probabilities of events nor a common causal explanation of the correlations between these events---provided that both events and common causes are interpreted non-classically. The violation of these Bell's inequalities is used for another purpose: it places a bound on the strength of correlations between these events.

\noi
I will proceed in the paper as follows. In Section 2 I analyze Bell's inequalities for classical probabilities, and in Section 3 for classical conditional probabilities. In Section 4 the two types will be compared. I turn to Bell's inequalities in terms of quantum probabilities in Section 5. In Section 6 I apply the results to the EPR-Bohm scenario and finally conclude in Section 7.

To make the notation simple, I drop both the hat and the asterisk from the next section on, that is I write $A$ instead of both $\hat{A}$ and also $A^*$. The semantics of $A$ will be clear from the context: in classical conditional probabilities $A$ will refer to a measurement outcome, in classical unconditional probabilities it will refer to a property/event, and in quantum probabilities it will refer to a projection.

My approach strongly relies on László E. Szabó's (2008) distinction between Bell's original inequalities and what he calls the Bell-Pitowsky inequalities. For a somewhat parallel research see Gömöri and Placek (2017).

\section{Case 1: Bell's inequalities for classical probabilities}

Consider real numbers $p_i$ and $p_{ij}$ in $[0,1]$ such that $i=1\dots n$ and $(i,j)\in S$ where $S$ is a subset of the index pairs $\left\{ (i,j)\left|i<j;\, i,j=1\dots n\right.\right\}$. When can these numbers be probabilities of certain events and their conjunctions? More precisely: given the numbers $p_i$ and $p_{ij}$, is there a classical probability space $(\O, \S, p)$ with events $A_i$ and $A_i \wedge A_j$ in $\S$ such that 
\begin{eqnarray*}
p_i &=& p(A_i) \\
p_{ij} &=& p(A_i \wedge A_j)
\end{eqnarray*}
In brief, do the numbers $p_i$ and $p_{ij}$ admit a \textit{Kolmogorovian representation}? 

Itamar Pitowsky's (1989) provided an elegant geometrical answer to this question. Arrange the numbers $p_i$ and $p_{ij}$ into a vector 
\begin{eqnarray*}
\vec{p}=\left(p_1, \dots, p_n; \dots,  p_{ij}, \dots\right)
\end{eqnarray*}
called \textit{correlation vector}. $\vec{p}$ is an element of an $n+\left|S\right|$ dimensional real linear space, $R(n,S)\cong\mathbb{R}^{n+\left|S\right|}$, where $\left|S\right|$ is the cardinality of $S$. 

Now, we construct a polytope in $R(n,S)$. Let $\varepsilon\in\left\{ 0,1\right\} ^{n}$. To each $\varepsilon$ assign a \textit{classical vertex vector} (truth-value function):  $\vec{u}^{\, \varepsilon}\in R(n,S)$ such that 
\begin{eqnarray*}
\vec{u}_{i}^{\,\varepsilon} & = & \varepsilon_{i}  \quad \quad i=1\ldots n\\
\vec{u}_{ij}^{\,\varepsilon} & = & \varepsilon_{i}\varepsilon_{j} \quad (i,j)\in S
\end{eqnarray*}
Then we define the \textit{classical correlation polytope} in $R(n,S)$ as the convex hull of the classical vertex vectors:
\begin{eqnarray*}
c(n,S) := \left\{\vec{p}\in R(n,S)\left|\vec{p}=\sum_{\varepsilon\in\left\{ 0,1\right\} ^{n}}\lambda_{\varepsilon}\vec{u}^{\, \varepsilon}\right.;\,\lambda_{\varepsilon}\geqslant 0;\,\sum_{\varepsilon\in\left\{ 0,1\right\} ^{n}}\lambda_{\varepsilon}=1\right\} 
\end{eqnarray*}
The polytope $c(n,S)$ is a simplex, that is any correlation vector in $c(n,S)$ has a unique expansion by classical vertex vectors.

Pitowsky's theorem (Pitowsky, 1989, p.\ 22) states that $\vec{p}$ admits a Kolmogorovian representation if and only if $\vec{p}\in c(n,S)$. That is numbers can be probabilities of certain events and their conjunctions if and only if the correlation vector composed of these numbers is in the classical correlation polytope.

Now, Bell's inequalities enter the scene as the \textit{facet inequalities} of the classical correlation polytopes. The simplest such correlation polytope is the one with $n=2$ and $S=\left\{ (1,2)\right\}$. In this case the vertices are: 
$(0,0;0)$, $(1,0;0)$, $(0,1;0)$ and $(1,1;1)$ and the facet inequalities (Bell's inequalities) are the following:
\begin{eqnarray*}
0\leqslant p_{12}\leqslant p_{1}, p_{2} \leqslant 1\\
p_{1}+p_{2}-p_{12}\leqslant 1
\end{eqnarray*}

Another famous polytope is $c(n,S)$ with $n=4$ and $S=\left\{ (1,3),(1,4),(2,3),(2,4)\right\} $. The facet inequalities are then the following: 
\begin{eqnarray}
0\leqslant p_{ij}\leqslant p_{i}, p_{j}\leqslant1 & \quad & i=1,2; \,\, j=3,4\\
p_{i}+p_{j}-p_{ij}\leqslant1&\quad & i=1,2; \,\, j=3,4\\
-1 \leqslant  p_{ij}+p_{i'j}+p_{ij'}-p_{i'j'}-p_{i}-p_{j} \leqslant 0&\quad & i,i'=1,2;\,\, j,j'=3,4;\,\, i\neq i';\,\, j\neq j' \quad  \label{CH}
\end{eqnarray}
The facet inequalities (\ref{CH}) are called the \textit{Clauser-Horne inequalities}. They express whether 4 + 4 real numbers can be regarded as the probability of four classical events and their certain conjunctions. 

\noi
A special type of the correlation vectors are the \textit{independence vectors} that is correlation vectors such that for all $(i,j)\in S$, $p_{ij} = p_ip_j$. It is easy to see that all independence vectors lie in the classical correlation polytope with coefficients
\begin{eqnarray} \label{lambda}
\lambda_{\varepsilon}= \prod_{i=1}^n p^*_i, \quad \quad \mbox{where} \, \, p^*_i = \left\{ \begin{array}{ll} p_i & \mbox{if}\ \varepsilon_i = 1 \\ 1-p_i & \mbox{if}\ \varepsilon_i = 0 \end{array} \right. 
\end{eqnarray}

Classical vertex vectors are independence vectors by definition; they are extremal points of the classical correlation polytope. For a classical vertex vector $p_i\in\{0,1\}$ for all $i =1 \dots n$. Thus we will sometimes call a classical vertex vector a \textit{deterministic} independence vector and an independence vector which is \textit{not} a classical vertex vector an \textit{indeterministic} independence vector. 

Although correlation vectors in $c(n,S)$ has a unique convex expansion by classical vertex vectors, they (if not classical vertex vectors) can have many convex expansions by indeterministic independence vectors. Moreover, for a classical correlation vector $\vec{p} \in c(n,S)$ (which is not a classical vertex vector) there always exist many sets of indeterministic independence vectors $\{\vec{p}^{\, \varepsilon}\}$ such that 
\begin{eqnarray*}
\vec{p}=\sum_\varepsilon \lambda_{\varepsilon}\vec{u}^{\, \varepsilon}=\sum_\varepsilon \lambda_{\varepsilon} \, \vec{p}^{\, \varepsilon}
\end{eqnarray*}
that is the coefficients $\lambda_{\varepsilon}$ of the expansion of $\vec{p}$ by classical vertex vectors and by indeterministic independence vectors are the same.\footnote{For example:
\begin{eqnarray}
\vec{p} &=& \left(\frac{2}{5}, \frac{2}{5}; \frac{1}{5} \right) \nonumber \\
&=& \frac{1}{5}\left(1,1;1\right) + \frac{1}{5}\left(1,0;0\right) + \frac{1}{5}\left(0,1;0\right) + \frac{2}{5}\left(0,0;0\right)  \nonumber \\
&=& \frac{1}{5}\left(\frac{1}{4}, \frac{1}{4}; \frac{1}{16} \right) + \frac{1}{5}\left(\frac{3+\sqrt{5}}{8}, \frac{3+\sqrt{5}}{8}; \frac{7+3\sqrt{5}}{32}\right) \nonumber \\
&& + \frac{1}{5}\left(\frac{3-\sqrt{5}}{8}, \frac{3-\sqrt{5}}{8}; \frac{7-3\sqrt{5}}{32} \right) + \frac{2}{5}\left(\frac{1}{2}, \frac{1}{2}; \frac{1}{4} \right)  \label{pelda}
\end{eqnarray}
}

Let us call a correlation vector which is not an independence vector a \textit{proper correlation vector}. For a proper correlation vector $\vec{p}$ there is at least one pair $(i,j)\in S$ such that $p_{ij}\neq p_ip_j$. When lying in the classical correlation polytope $\vec{p}$ represents the probabilities of certain events and their conjunctions and the events associated to indices $i$ and $j$ are correlated:
\begin{eqnarray}
p(A_{i}\w A_{j}) \neq p(A_{i}) \, p(A_{j})\label{corr} 
\end{eqnarray}
Now, we ask the following question: When do the correlations in the Kolmogorovian representation of a proper correlation vector in $c(n,S)$ have a common causal explanation?

A common cause in the Reichenbachian sense (Reichenbach, 1956) is a screener-off partition of the algebra $\S$ (or an extension of the algebra; see Hofer-Szabó, Rédei, Szabó, 2013, Ch.\ 3 and 6). In other words, correlations (\ref{corr}) are said to have a \textit{joint common causal explanation} if there is a partition $\{C_k\}$ ($k\in K$) of $\S$ such that for any $(i,j)$ in $S$ and $k\in K$
\begin{eqnarray}
p(A_i \wedge A_j| C_k ) &=& p(A_i | C_k) \, p(A_j | C_k) \label{screen} 
\end{eqnarray}

Introduce the notation
\begin{eqnarray*}
p_i^k &=& p(A_i|C_k) \\
c_k &=& p(C_k) 
\end{eqnarray*}
where $\sum_k c_k = 1$ and construct for each $k\in K$ a \textit{common cause vector}
\begin{eqnarray*}
\vec{p}^{\,k}=\left(p_1^k, \dots, p_n^k; \dots,  p_i^kp_j^k, \dots\right) 
\end{eqnarray*}
for the classical correlation vector $\vec{p}$. Due to (\ref{screen}) and the theorem of total probability
\begin{eqnarray*}
\vec{p} = \sum_k c_k \, \vec{p}^{\,k} 
\end{eqnarray*}
Since the common cause vectors are independence vectors lying in the classical correlation polytope $c(n,S)$, therefore their convex combination $\vec{p}$ also lies in the classical correlation polytope---which, of course,  we knew since we assumed that $\vec{p}$ has a Kolmogorovian representation.

We call a common cause vector $\vec{p}^{\,k}$ \textit{deterministic}, if $\vec{p}^{\,k}$ is a deterministic independence vector (classical vertex vector); otherwise we call it \textit{indeterministic}. All classical correlation vectors have $2^n$ deterministic common cause vectors, namely the classical vertex vectors. In this cases $k = \varepsilon \in \left\{ 0,1\right\} ^{n}$ and the probabilities are:
\begin{eqnarray*}
p_i^\varepsilon &=& \varepsilon_i \\
c_\varepsilon &=& \lambda_{\varepsilon} 
\end{eqnarray*}
where $\lambda_{\varepsilon}$ is specified in (\ref{lambda}). Classical correlation vectors which are not classical vertex vectors also can be expanded as a convex sum of indeterministic common cause vectors in many different ways.

Conversely, knowing the $k$ common cause vectors and the probabilities $c_k$ alone, one can easily construct a common causal explanation of the correlations (\ref{corr}). First, construct the classical probability space $\S$ associated to the correlation vector $\vec{p}$ (for the details see Pitowsky 1989, p.\ 23). Then extend $\S$ containing the events $A_i$ and $A_i \wedge A_j$ such that the extended probability space contains also the common causes $\{C_k\}$ (for such an extension see Hofer-Szabó, Rédei, Szabó, 1999).

\noi
To sum up, in the case of classical probabilities the fulfillment of Bell's inequalities is a necessary and sufficient condition for a set of numbers to be the probability of certain events and their conjunctions. If these events are correlated, then the correlations will always have a common causal explanation. In other words, having a common causal explanation does not put a further constraint on the correlation vectors. Thus, Bell's inequalities have a double meaning: they test whether numbers can represent probabilities of events and at the same time whether the correlations between these events (provided they exist) have a joint common cause. These two meanings of Bell's inequalities  will split up in the next section where we treat Bell's inequalities with classical \textit{conditional} probabilities.

\section{Case 2: Bell's inequalities for classical conditional probabilities}

Just as in the previous section suppose that we are given real numbers $p_i$ and $p_{ij}$ in $[0,1]$ with $i=1\dots n$ and $(i,j)\in S$. But now we ask: when can these numbers be \textit{conditional} probabilities of certain events and their conjunctions? More precisely: given the numbers $p_i$ and $p_{ij}$, do there exist events $A_i$ and $a_i$ ($i=1\dots n$) in a classical probability space $(\O, \S, p)$ such that $p_i$ and $p_{ij}$ are the following classical conditional probabilities:
\begin{eqnarray*}
p_i &=& p(A_i|a_i) \label{pAa}\\
p_{ij} &=& p(A_i \wedge A_j | a_i \wedge a_j) \label{pAAaa}
\end{eqnarray*}
Or again in brief, do the numbers $p_i$ and $p_{ij}$ admit a \textit{conditional Kolmogorovian representation}?

The answer here is more permissive. Except for some extremal values the numbers $p_i$ and $p_{ij}$ always admit a conditional Kolmogorovian representation: any correlation vector $\vec{p}$ admits a conditional Kolmogorovian representation if $\vec{p}$ has no $(i,j)\in S$ such that either (i) $p_i=0$ or $p_j=0$ but $p_{ij} \neq 0$; or (ii) $p_i=p_j=1$ but $p_{ij} \neq 1$. 

Obviously, any Kolmogorovian representations is a conditional Kolmogorovian representation with $a_i = \O$ for all $i=1\dots n$. However, correlation vectors admitting a conditional Kolmogorovian representation are not necessarily in the classical correlation polytope.\footnote{For example consider the following correlation vector in $R(n,S)$ with $n=2$ and $S = \{1,2\}$:
\begin{eqnarray*}
\vec{p}=\left(\frac{2}{3}, \frac{2}{3}; \frac{1}{5} \right)
\end{eqnarray*}
The vector $\vec{p}$ violates Bell's inequality 
\begin{eqnarray*}
p_{1}+p_{2}-p_{12}\leqslant 1
\end{eqnarray*}
hence it is not in $c(n,S)$ and, consequently, it does not admit a Kolmogorovian representation. However, $\vec{p}$ admits a conditional Kolmogorovian representation with the following atomic events and probabilities:
\begin{eqnarray*}
p(A_1 \w A_2 \w a_1 \w a_2) &=& \frac{1}{25} \\
p(A^\perp_1 \w A^\perp_2 \w a_1 \w a_2) &=& \frac{4}{25} \\
p(A^\perp_1 \w A_2 \w a^\perp_1 \w a_2) = p(A_1 \w A^\perp_2 \w a_1 \w a^\perp_2) &=& \frac{9}{25} \\
p(A^\perp_1 \w A^\perp_2 \w a^\perp_1 \w a_2) = p(A^\perp_1 \w A^\perp_2 \w a_1 \w a^\perp_2) &=& \frac{1}{25}
\end{eqnarray*}
(The probability of all other atomic events is $0$.)}

Now, any correlation vector can be expressed as a convex combination of (not necessarily classical) vertex vectors. A \textit{vertex vector}  $\vec{u}$ is defined as follows: $\vec{u}_i, \vec{u}_{ij} \in \{0,1\}$ for all $i=1\dots n$ and $(i,j)\in S$. Obviously, classical vertex vectors are vertex vectors but not every vertex vector is classical. For example the vertex vectors $(0,0;1)$, $(1,0;1)$, $(0,1;1)$ or $(1,1;0)$ in $R(n,S)$ with $n=2$ and $S = \{1,2\}$ are not classical. There are $2^{n+|S|}$ different vertex vectors $\vec{u}^{\, k}$ in $R(n,S)$ and $2^n$ different classical vertex vectors $\vec{u}^{\, \varepsilon}$.

Denote the convex hull of the vertex vectors by
\begin{eqnarray*}
u(n,S) := \left\{\vec{v}\in R(n,S)\left|\vec{v}=\sum_{k =1}^{2^{n+|S|}}\lambda_{k}\vec{u}^{\,k}\right.;\,\lambda_{k}\geqslant 0;\,\sum_{k =1}^{2^{n+|S|}}\lambda_{k}=1 \right\} 
\end{eqnarray*}
Contrary to $c(n,S)$, the polytope $u(n,S)$ is not a simplex, hence the expansion of the correlation vectors in $u(n,S)$ by vertex vectors is typically not unique.\footnote{The correlation vector 
\begin{eqnarray*}
\vec{p}=\left(\frac{2}{3}, \frac{2}{3}; \frac{1}{5} \right)
\end{eqnarray*}
for example can be expanded in many different ways: 
\begin{eqnarray*}
\vec{p} &=& \frac{1}{3} (0,1;0) + \frac{1}{3} (1,0;0) + \frac{1}{5} (1,1;1) + \frac{2}{15} (1,1;0) \\
&=& \frac{1}{3} (0,0;0) + \frac{1}{5} (1,1;1) + \frac{7}{15} (1,1;0)
\end{eqnarray*}
etc.}

Now, the set of correlation vectors admitting a conditional Kolmogorovian representation is dense in $u(n,S)$: all interior points of $u(n,S)$ admit a conditional Kolmogorovian representation and also all surface points, except for which there is a pair $(i,j)\in S$ such that either (i) $p_i=0$ or $p_j=0$ but $p_{ij} \neq 0$; or (ii) $p_i=p_j=1$ but $p_{ij} \neq 1$. Denote the set of correlation vectors admitting a conditional Kolmogorovian representation by $u'(n,S)$.
 
\noi
Now, let's go over to the common causal explanation of conditional correlations. Let $\vec{p}$ be a proper correlation vector in $u'(n,S)$. Lying in $u'(n,S)$ the correlation vector $\vec{p}$ represents the conditional probabilities of certain events and their conjunctions and the events associated to some pairs $(i,j)\in S$ are \textit{conditionally correlated}:
\begin{eqnarray}
p(A_{i}\w A_{j}|a_i \w a_j) \neq p(A_{i}|a_i) \, p(A_{j}|a_j)\label{condcorr} 
\end{eqnarray}

Interpret now the events $A_i$ and $a_i$ in the context of physical experiments: Let $a_i$ represent a possible \textit{measurement} that an experimenter can perform on an object. Then the event $a_i \w a_j$ will represent the joint performance of measurements $a_i$ and $a_j$. Let furthermore the event $A_i$ represent an outcome of measurement $a_i$ and $A_i \w A_j$ represent an outcome of the jointly performed measurement $a_i \w a_j$. Then (\ref{condcorr}) expresses a correlation between two measurement outcomes provided their measurements have been performed.

When do the conditional correlations in a conditional Kolmogorovian representation of $\vec{p}$ have a common causal explanation?

The set of conditional correlations are said to have a \textit{non-conspiratorial joint common causal explanation} if there is a partition $\{C_k\}$ ($k\in K$) of $\S$ such that for any $(i,j)$ in $S$ and $k\in K$
\begin{eqnarray}
p(A_i \wedge A_j|a_{i}\wedge a_{j} \wedge C_k ) &=& p(A_i |a_i \wedge C_k) \, p(A_j | a_j \wedge C_k) \label{condscreen} \\
p(a_{i}\wedge a_{j}\wedge C_{k}) &=& p(a_{i} \wedge a_{j}) \, p(C_{k}) \label{nocons}
\end{eqnarray}
Equations (\ref{condscreen}) express the Reichenbachian idea that the common cause is to \textit{screen off} all correlations. Equations (\ref{nocons}) express the so-called \textit{no-conspiracy}, the idea that common causes should be causally, and hence probabilistically, independent of the measurement choices. The common causal explanation is \textit{joint} since all correlations (\ref{condcorr}) have the same common cause. 

Now, suppose the correlations in the a given conditional Kolmogorovian representation of (\ref{condcorr}) of $\vec{p}$ in $u'(n,S)$ have a non-conspiratorial joint common causal explanation. Introduce again the notation
\begin{eqnarray*}
p_i^k &=& p(A_i|a_i \w C_k) \\
c_k &=& p(C_k) 
\end{eqnarray*}
and consider the $k$ common cause vectors of the correlation vector $\vec{p}$:
\begin{eqnarray*}
\vec{p}^{\,k}=\left(p_1^k, \dots, p_n^k; \dots,  p_i^kp_j^k, \dots\right)
\end{eqnarray*}
We call a non-conspiratorial joint common cause \textit{deterministic} if $\vec{p}^{\,k}$ is a deterministic common cause vector for all $k\in K$; otherwise we call it \textit{indeterministic}. We call a deterministic non-conspiratorial joint common cause $\{C_k\}$ ($k\in K$) a \textit{property}; and an indeterministic common cause a \textit{propensity}.

Now, due to (\ref{condscreen})-(\ref{nocons}) and the theorem of total probability
\begin{eqnarray*}
\vec{p} = \sum_k c_k \, \vec{p}^{\,k} 
\end{eqnarray*}
Since common cause vectors are independence vectors lying in $c(n,S)$, their convex combination also lies in the classical correlation polytope $c(n,S)$. Thus, $\vec{p}$ being a classical correlation vector is a necessary condition for a conditional Kolmogorovian representation of $\vec{p}$ to have a non-conspiratorial joint common causal explanation.

Conversely, knowing the $k$ common cause vectors and the probabilities $c_k$ alone, one can construct the classical probability space $\S$ with the conditionally correlating events $A_i$ and $a_i$ and the common causes $\{C_k\}$. 

\noi
Observe that the situation is now different from the one in the previous section: the numbers $p_i$ and $p_{ij}$ can be conditional probabilities of events and their conjunctions even if they violate the corresponding Bell inequalities. However, the conditional correlations between these events have a non-conspiratorial joint common causal explanation if and only if the correlation vector composed of the numbers $p_i$ and $p_{ij}$ lies in the classical correlation polytope. In other words, in case of classical \textit{conditional} probabilities Bell's inequalities do not test the conditional Kolmogorovian representability but whether correlations can be given a common causal explanation.

\section{Relating Case 1 and Case 2}

How do the Kolmogorovian and the conditional Kolmogorovian representation relate to one another?

In this section I will show that (i) a conditional Kolmogorovian representation of a classical correlation vector has a property explanation only if the representation is non-signaling (see below); and (ii) a correlation vector $\vec{p}$ has a Kolmogorovian representation if and only if it has a property explanation for any non-signaling  conditional Kolmogorovian representation.

Let $\vec{p}$ be a proper correlation vector in $c(n,S)$ and consider a conditional Kolmogorovian representation of $\vec{p}$. Obviously, there are many such representations of $\vec{p}$ depending on the measurement conditions $a_i$. Let's say, somewhat loosely, that a conditional Kolmogorovian representation has a property/propensity explanation if the conditional correlations in the representation have a property/propensity explanation. 

First, we claim that a conditional Kolmogorovian representation of a correlation vector $\vec{p}$ in $c(n,S)$ has a property explanation only if the representation satisfies \textit{non-signaling}:
\begin{eqnarray}
p(A_i|a_i) &=& p(A_i|a_i \wedge a_j) \label{nosign1} \\
p(A_j|a_j) &=& p(A_j|a_i \wedge a_j) \label{nosign2} 
\end{eqnarray}
for any $(i,j) \in S$. Note that non-signaling is not a feature of the correlation vector itself but of the representation. For the same correlation vector in $c(n,S)$ one can provide both non-signaling and also signaling representations.

Now, a conditional Kolmogorovian representation can have a property explanation only if it satisfies non-signaling. Recall namely that for a property $\{C_k\}$: 
\begin{eqnarray*}
p(A_i | a_i \w C_k) \in \{0,1\}
\end{eqnarray*}
and hence 
\begin{eqnarray} 
p(A_i | a_i \w C_k) = p(A_i | a_i \w a_j \w C_k) \label{hiddennosign}
\end{eqnarray}
for any $i,j =1 \dots n$ and $k \in K$. But (\ref{hiddennosign}) together with no-conspiracy (\ref{nocons}) and the theorem of total probability imply non-signaling (\ref{nosign1})-(\ref{nosign2}). Thus, satisfying non-signaling is a necessary condition for a conditional Kolmogorovian representation to have a property explanation. Signaling conditional Kolmogorovian representations do not have a property explanation.

Second, suppose that a conditional Kolmogorovian representation of $\vec{p}$ have a property explanation. That is $\vec{p}$ has a conditional Kolmogorovian representation in a classical probability space $\S$ and all the conditionally correlating event pairs have a non-conspiratorial deterministic joint common cause in $\S$. Then these properties (deterministic common causes) provide a Kolmogorovian (unconditional) representation for $\vec{p}$. 

Observe namely that
\begin{eqnarray*}
p_i &=& p(A_i|a_i) = \frac{p(A_i \wedge a_i)}{p(a_i)} = \frac{\sum_k p(A_i \wedge a_i \wedge C_k)}{p(a_i)} = \frac{\sum_k p(A_i|a_i \wedge C_k) p(a_i \wedge C_k)}{p(a_i)} \nonumber \\
&\stackrel{*}{=}& \frac{\sum_k p(A_i|a_i \wedge C_k) p(a_i) p(C_k)}{p(a_i)} = \sum_k p(A_i|a_i \wedge C_k) p(C_k) = \sum_{k:\, p_i^k=1} p(C_k)
\end{eqnarray*}
where the equation $\stackrel{*}{=}$ holds due to no-conspiracy (\ref{nocons}) and the symbol $\sum_{k:\, p_i^k=1}$ means that we sum up for all $k$ for which $p_i^k=1$. Similarly, using (\ref{condscreen})-(\ref{nocons}) one obtains 
\begin{eqnarray*}
p_{ij} = \sum_{k:\, p_i^k=1,\, \, p_j^k=1} p(C_k)
\end{eqnarray*}
That is the events 
\begin{eqnarray*}
C_i &=& \vee_{k:\, p_i^k=1} C_k \\
C_i \wedge C_j &=& \vee_{k:\, p_i^k=1,\, \, p_j^k=1} C_k 
\end{eqnarray*}
provide a Kolmogorovian representation for the numbers $p_i$ and $p_{ij}$. 

Third, conversely, if $\vec{p}$ admits a Kolmogorovian representation, then it also admits a property explanation for any non-signaling conditional Kolmogorovian representation. 

Namely, if $\vec{p}$ admits a Kolmogorovian representation, then there is a partition $\{C_\varepsilon\}$ in $\S$ such that
\begin{eqnarray}
p_i &=& p(C_i) = \sum_{\varepsilon: \, \varepsilon_i =i} p(C_\varepsilon) = \sum_{\varepsilon: \, \varepsilon_i =i} \l_\varepsilon \label{pC} \\
p_{ij} &=& p(C_i \wedge C_j) = \sum_{\varepsilon: \, \varepsilon_i =1, \,\varepsilon_j =1} p(C_\varepsilon) = \sum_{\varepsilon: \, \varepsilon_i =1, \,\varepsilon_j =1} \l_\varepsilon \label{pCC}
\end{eqnarray}
Now, consider a non-signaling conditional Kolmogorovian representation of $\vec{p}$, that is let there be events $A_i$ and $a_i$ in $\S'$ such that
\begin{eqnarray*}
p_i &=& p(A_i|a_i) \\
p_{ij} &=& p(A_i \wedge A_j | a_i \wedge a_j) 
\end{eqnarray*}
and suppose that non-signaling (\ref{nosign1})-(\ref{nosign2}) hold for any $(i,j) \in S$.

Now, the events $C_\varepsilon$ provide a propensity explanation for the conditional Kolmogorovian representation of $\vec{p}$ in the following sense. First, let $\varepsilon, \varepsilon' \in \{0,1\}^n$ and define the events $a_{\varepsilon'}$ in $\S'$ as follows: 
\begin{eqnarray*}
a_{\varepsilon'} := \wedge_i \, a_{\varepsilon'_i} \quad \quad \mbox{where}  \, \, a_{\varepsilon'_i} =\left\{ \begin{array}{ll} a_i & \mbox{if}\ \varepsilon'_i =1 \\ \overline{a}_i & \mbox{if}\ \varepsilon'_i =0 \end{array} \right.
\end{eqnarray*}
Then, extend the algebra $\S'$ to $\S''$ generated by the atomic events $D_{\varepsilon' \! , \varepsilon}$ defined as follows:
\begin{eqnarray}
\vee_{\varepsilon} D_{\varepsilon' \! , \varepsilon} &=& a_{\varepsilon'} \label{1} \\
\vee_{\varepsilon'} D_{\varepsilon' \! , \varepsilon} &=& C_{\varepsilon} \\
p(D_{\varepsilon' \! , \varepsilon}) &=& p(a_{\varepsilon'}) \, p(C_{\varepsilon}) = p(a_{\varepsilon'}) \, \l_{\varepsilon} \\
p(A_i | D_{\varepsilon' \! , \varepsilon}) &=& \varepsilon'_i \, p^\varepsilon_i \\
p(A_i \wedge A_j | D_{\varepsilon' \! , \varepsilon}) &=& \varepsilon'_i  \, p^\varepsilon_i \varepsilon'_j \, p^\varepsilon_j \label{5} 
\end{eqnarray}
where the numbers $p^\varepsilon_i \in [0,1]$ will be specified below. Now, using (\ref{1})-(\ref{5}) one obtains
\begin{eqnarray*}
p(A_i \wedge A_j  |  a_i \wedge a_j  \w C_{\varepsilon}) = p^\varepsilon_i \, p^\varepsilon_j  = p(A_i | a_i \w C_{\varepsilon}) \, p(A_j | a_j \w C_{\varepsilon}) \\
p(a_{i}\wedge a_{j}\wedge C_{\varepsilon}) = \sum_{\varepsilon' : \, \varepsilon'_i=1, \, \varepsilon'_j=1} p(D_{\varepsilon' \! , \varepsilon}) = p(a_{i} \wedge a_{j}) \, p(C_{\varepsilon}) 
\end{eqnarray*}
That is the partition $\{C_{\varepsilon}\}$ provides a propensity explanation for the conditional Kolmogorovian representation of $\vec{p}$ with probabilities specified in (\ref{pC})-(\ref{pCC}). Now
\begin{eqnarray*}
p(A_i | a_{\varepsilon'}) &=& \frac{p(A_i \w a_{\varepsilon'})}{p(a_{\varepsilon'})}  =  \frac{\sum_{\varepsilon} p(A_i \w D_{\varepsilon' \! , \varepsilon})}{p(a_{\varepsilon'})} = \frac{\sum_{\varepsilon} p(A_i | D_{\varepsilon' \! , \varepsilon}) \, p(D_{\varepsilon' \! , \varepsilon})}{p(a_{\varepsilon'})} \nonumber \\
&=& \frac{\sum_{\varepsilon} \varepsilon'_i \, p^\varepsilon_i \, p(a_{\varepsilon'}) \l_{\varepsilon}}{p(a_{\varepsilon'})} = \varepsilon'_i \sum_{\varepsilon}  p^\varepsilon_i \, \l_{\varepsilon}
\end{eqnarray*}
that is 
\begin{eqnarray} \label{p_i}
p_i = p(A_i | a_i) = \sum_{\varepsilon}  p^\varepsilon_i \, \l_{\varepsilon}
\end{eqnarray}
and similarly
\begin{eqnarray} \label{p_ip_j}
p_{ij} = p(A_i \w A_j | a_i \w a_j) = \sum_{\varepsilon}  p^\varepsilon_i \, p^\varepsilon_j \, \l_{\varepsilon}
\end{eqnarray}
Composing $2^n$ independence vectors $\vec{p}^{\, \varepsilon}$ from the numbers $p^\varepsilon_i$: 
\begin{eqnarray*}
\vec{p}^{\, \varepsilon} =\left(\dots p^\varepsilon_i \dots p^\varepsilon_j \dots ; \dots p^\varepsilon_i p^\varepsilon_j \dots\right)
\end{eqnarray*}
(\ref{p_i})-(\ref{p_ip_j}) reads as follows: 
\begin{eqnarray*}
\vec{p}= \sum_\varepsilon \l_\varepsilon \, \vec{p}^{\, \varepsilon}
\end{eqnarray*}

This means that the numbers $p^\varepsilon_i$ are to be taken from $[0,1]$ such that the $2^n$ independence vectors $\vec{p}^{\, \varepsilon}$ provide a convex combination for $\vec{p}$ with the \textit{same} coefficients $\l_\varepsilon$ as the vertex vectors $\vec{u}^{\, \varepsilon}$ do. One such expansion always exists. Namely, when $\vec{p}^{\, \varepsilon}= \vec{u}^{\, \varepsilon}$. In this case $p^\varepsilon_i = \varepsilon_i$ for any $i=1 \dots n$ and $\varepsilon \in \{0,1\}^n$ and (\ref{p_i})-(\ref{p_ip_j}) reads as follows: 
\begin{eqnarray*} 
p_i &=& p(A_i | a_i) = \sum_{\varepsilon} \varepsilon_i\, p(C_\varepsilon) = p(C_i) \\
p_{ij} &=& p(A_i \w A_j | a_i \w a_j) = \sum_{\varepsilon} \varepsilon_i \varepsilon_j\, p(C_\varepsilon) = p(C_i \w C_j) 
\end{eqnarray*}
That is we obtain a property explanation for the conditional Kolmogorovian representation of $\vec{p}$. However, for classical correlation vectors which are not classical vertex vectors one can also provide for $\vec{p}$ various convex combinations by indeterministic independence vectors with the coefficients $\l_\varepsilon$ (see (\ref{pelda}) as an example). In this case we obtain a propensity explanation for the given conditional Kolmogorovian representation of $\vec{p}$. 

To sum up, a correlation vector $\vec{p}$ has a Kolmogorovian representation if and only if it has a property explanation for any non-signaling  conditional Kolmogorovian representation. This equivalence justifies retrospectively why we used the term ``property'' both for a deterministic common cause in the conditional Kolmogorovian representation and also as a synonym of ``event'' in the unconditional Kolmogorovian representation.

\section{Case 3: Bell's inequalities for quantum probabilities}

Again start with real numbers $p_i$ and $p_{ij}$ in $[0,1]$ with $i=1\dots n$ and $(i,j)\in S$. But now the question is the following: when can these numbers be \textit{quantum} probabilities? That is, given the numbers $p_i$ and $p_{ij}$, do there exist projections $A_i$ ($i=1\dots n$) representing quantum events in a quantum probability space $(\cP(H), \rho)$, where $\cP(H)$ is the projection lattice of a Hilbert space $H$ and $\rho$ is a density operator   on $H$ representing the quantum state, such that $p_i$ and $p_{ij}$ are the following quantum probabilities:
\begin{eqnarray*}
p_i &=& \mbox{Tr}(\rho A_i) \label{TrA}\\
p_{ij} &=& \mbox{Tr}(\rho (A_i \w A_j)) \label{TrAA}
\end{eqnarray*}
(Here $A_i \w A_j$ denotes the projection projecting on the intersection of the closed subspaces of $H$ onto which $A_i$ and $A_j$ are projecting.)  Again in brief, do the numbers $p_i$ and $p_{ij}$ admit a \textit{quantum representation}?

The answer is again given by Pitowsky (1989, p.\ 72). Introduce the notion of a quantum vertex vector. A quantum vertex vector is a vertex vector in $R(n,S)$ such that 
\begin{eqnarray*}
\vec{u}_{i}^{\,\varepsilon} & = & \varepsilon_{i}  \quad \quad i=1\ldots n\\
\vec{u}_{ij}^{\,\varepsilon} & \leqslant & \varepsilon_{i} \varepsilon_{j} \quad (i,j)\in S
\end{eqnarray*}
Obvious, any classical vertex vector is a quantum vertex vector and any quantum vertex vector is a vertex vector, but the reverse inclusion does not hold. In $R(n,S)$ with $n=2$ and $S = \{1,2\}$ for example the vertex vector $(1,1;0)$ is quantum but not classical, and the vertex vectors $(0,0;1)$, $(1,0;1)$, $(0,1;1)$ are not even quantum. 

Denote by $q(n,S)$ the convex hull of quantum vertex vectors. Pitowsky then shows that almost all correlation vectors in $q(n,S)$ admit a quantum representation. More precisely, Pitowsky shows that---denoting by $q'(n,S)$ the set of quantum correlation vectors that is the set of those correlation vectors which admit a quantum representation---the following holds: 
\begin{itemize}
\item[(i)] $c(n,S) \subset q'(n,S) \subset q(n,S)$;
\item[(ii)] $q'(n,S)$ is convex (but not closed);
\item[(iii)] $q'(n,S)$ contains the interior of $q(n,S)$
\end{itemize}
We can add to this our result in the previous section:
\begin{itemize}
\item[(iv)] $q(n,S) \subset u'(n,S) \subset u(n,S)$
\end{itemize}

Thus, the set of numbers admitting a quantum representation is strictly larger than the set of numbers admitting a Kolmogorovian representation but strictly smaller than the set of numbers admitting a conditional Kolmogorovian representation.

\noi
Let us turn now to the question of the common causal explanation. Let $\vec{p}$ be a proper quantum correlation vector. $\vec{p}$ then represents the quantum probabilities of certain quantum events and their conjunctions and the events associated to some pairs $(i,j)\in S$ are \textit{correlated}:
\begin{eqnarray}
\mbox{Tr}(\rho(A_{i}\w A_{j})) \neq \mbox{Tr}(\rho A_{i})\, \mbox{Tr}(\rho A_{j}) \label{quantcorr} 
\end{eqnarray}

When do the correlations (\ref{quantcorr}) have a common causal explanation?

The set of quantum correlations has a \textit{joint quantum common causal explanation} if there is a partition $\{C_k\}$ ($k\in K$) in $\cP(H)$ (that is a set of mutually orthogonal projection adding up to the unity) such that for any $(i,j)$ in $S$ and $k\in K$
\begin{eqnarray}
\mbox{Tr}(\rho_k(A_{i}\w A_{j}))  &=& \mbox{Tr}(\rho_kA_{i}) \, \mbox{Tr}(\rho_kA_{j}) \label{quantscreen} 
\end{eqnarray}
where 
\begin{eqnarray*}
\rho_k := \frac{C_k\rho \, C_k}{\mbox{Tr}(\rho \, C_k)} 
\end{eqnarray*}
is the density operator $\rho$ after a selective measurement by $C_k$. If the partition $\{C_k\}$ is commuting with the each correlating pair $(A_i, A_j)$, then we call the common causes \textit{commuting}, otherwise \textit{noncommuting}.

Now, suppose the correlations (\ref{quantcorr}) have a joint quantum common causal explanation. Does it follow that $\vec{p}$ is in $c(n,S)$ that is Bell's inequalities are satisfied?

Introduce again the notation
\begin{eqnarray*}
p_i^k &=& \mbox{Tr}(\rho_kA_{i})\\
c_k &=& \mbox{Tr}(\rho \, C_k)
\end{eqnarray*}
and consider the $k$ common cause vectors of the correlation vector $\vec{p}$:
\begin{eqnarray*}
\vec{p}^{\,k}=\left(p_1^k, \dots, p_n^k; \dots,  p_i^kp_j^k, \dots\right)
\end{eqnarray*}
The common cause vectors are independence vectors. Hence, using (\ref{quantscreen}) and the theorem of total probability, the convex combination of the common cause vectors 
\begin{eqnarray}\label{constr}
\vec{p}^{\,c}= \sum_k c_k \, \vec{p}^{\,k} 
\end{eqnarray}
will be in $c(n,S)$. However $\vec{p}^{\,c}$ is not necessarily identical with the original correlation vector $\vec{p}$. More precisely, $\vec{p}^{\,c}=\vec{p}$ for any $\rho$ if $\{C_k\}$ are commuting common causes. But if $\{C_k\}$ are noncommuting common causes, then $\vec{p}^{\,c}$ and $\vec{p}$ can be different and hence even if $\vec{p}^{\,c} \in c(n,S)$, $\vec{p}$ might be outside $c(n,S)$. In short, a quantum correlation vector can have a joint \textit{noncommuting} common causal explanation even if it lies outside the classical correlation polytope. The correlation vector $\vec{p}$ is confined in $c(n,S)$ only if the common causes are required to be \textit{commuting}.\footnote{For the details and for a concrete example see Hofer-Szabó and Vecsernyés, 2013, 2018.}

\noi
To sum up, in the case of quantum probabilities we found a scenario different from both previous cases. Here Bell's inequalities neither put a constraint on whether numbers can be quantum probabilities nor on whether the correlation between events with the prescribed probability can have a common causal explanation. Bell's inequalities constrain common causal explanations only if common causes are understood as commuting common causes. 

Perhaps it is worth mentioning that in algebraic quantum field theory (Rédei and Summers, 2007; Hofer-Szabó and Vecsernyés, 2013) and quantum information theory (Bengtson and Zyczkowski, 2006) the violation of Bell's inequalities composed of quantum probabilities is used for another purpose: it places a bound on the strength of correlations between certain events. Abstractly, one starts with two mutually commuting $C^*$-subalgebras $\mathcal A$ and $\mathcal B$ of a $C^*$-algebra $\mathcal C$ and defines a \textit{Bell operator} $R$ for the pair ($\mathcal A, \mathcal B$) as an element of the following set:
\begin{eqnarray*}
\mathbb{B}(\mathcal A, \mathcal B) & := & \left\lbrace \frac{1}{2} \big( A_1(B_1 + B_2) + A_2(B_1 - B_2) \big) \, {\big |} \, A_i =A_i^* \in \mathcal A; \, B_i =B_i^* \in \mathcal B; \, -\UN \leqslant A_i, B_i  \leqslant \UN \right\rbrace 
\end{eqnarray*}
where $\UN$ is the unit element of $\mathcal C$. Then one can prove that for any Bell operator $R$, $| \phi(R) | \leqslant \sqrt{2}$ for any state $\phi$; but $|\phi(R)| \leqslant 1$ for separable states (i.e. for convex combinations of product states).

In other words, in these disciplines one fixes a Bell operator (a witness operator) and scrolls over the different quantum states to see which of them is separable. In this reading, Bell's inequalities neither test Kolmogorovian representation nor common causal explanation but separability of certain normalized positive linear functionals. Obviously, this role of Bell's inequalities is completely different from the one analyzed in this paper.

\section{The EPR-Bohm scenario}

Bell's inequalities in the EPR-Bohm scenario are standardly said to be violated. But which Bell inequalities and what does this violation mean?

Let us start in the reverse order, with quantum representation. Consider the EPR-Bohm scenario with pairs of spin-$\frac{1}{2}$ particles prepared in the singlet state. In quantum mechanics the state of the system and the quantum events are represented by matrices on $M_2 \otimes M_2$, where $M_2$ is the algebra of the two-dimensional complex matrices. The singlet state is represented as:
\begin{eqnarray*}
\rho^s = \frac{1}{4}\big(\UN \otimes \UN - \sum_{k=1}^3 \s_k \otimes \s_k\big)
\end{eqnarray*}
and the event that the spin of the particle is ''up'' on the left wing  in direction $\vec{a}_i$ ($i=1,2$); on the right wing in direction $\vec{b}_j$ ($i=3,4$); and on both wings in directions $\vec{a}_i$ and $\vec{b}_j$, respectively, are represented as:
\begin{eqnarray*}
A_i &=& \frac{1}{2}\big((\UN + \vec{a}_i\cdot\vec{\s}) \otimes \UN\big)\\
A_j &=& \frac{1}{2}\big( \UN \otimes (\UN + \vec{b}_j\cdot\vec{\s})\big) \\
A_{ij} &=& \frac{1}{4}\big((\UN + \vec{a}_i\cdot\vec{\s}) \otimes (\UN + \vec{b}_j\cdot\vec{\s})\big) 
\end{eqnarray*}
where $\UN$ is the identity matrix on $M_2$, $\vec{\s}=(\s_1, \s_2, \s_3)$ is the Pauli vector, and $\vec{a}_i$ ($i=1,2$) and $\vec{b}_j$ ($j=3,4$) are the spin measurement directions on the left and right wing, respectively. Furthermore, $A_{ij} = A_i A_j = A_i \w A_j$ since $A_i$ and $A_j$ are commuting.

The quantum probabilities are generated by the trace formula: 
\begin{eqnarray}
p_i &=& \mbox{Tr}(\rho^s A_i) = \frac{1}{2} \label{q1} \\\
p_{j} &=& \mbox{Tr}(\rho^s A_j) = \frac{1}{2}  \\
p_{ij} &=& \mbox{Tr}(\rho^s A_{ij})  = \frac{1}{2} \sin^2\left(\frac{\theta _{ij}}{2}\right) \label{q3} 
\end{eqnarray}
where $\theta _{ij}$ denotes the angle between directions $\vec{a}_i$ and $\vec{b}_j$. As it is well known, for the measurement directions 
\begin{eqnarray*}
\vec{a}_1 = (0,1,0) &\quad & \vec{b}_3 = \frac{1}{\sqrt{2}} (1,1,0) \label{a1} \\
\vec{a}_2 = (1,0,0) &\quad & \vec{b}_4 = \frac{1}{\sqrt{2}} (-1,1,0) \label{b2} 
\end{eqnarray*}
the Clauser-Horne inequality
\begin{eqnarray}
-1 \leqslant  p_{13}+p_{23}+p_{14}-p_{24}-p_{1}-p_{3} = - \frac{1+\sqrt{2}}{2} \label{CH'}
\end{eqnarray}
is violated. 

What does the violation of the Clauser-Horne inequality (\ref{CH'}) mean with respect to the quantum representation? As clarified in Section 5, it does not mean that the numbers (\ref{q1})-(\ref{q3}) cannot be given a quantum mechanical representation. Equations (\ref{q1})-(\ref{q3}) just provide one. On the other hand, the violation of (\ref{CH'}) neither means that the EPR correlations
\begin{eqnarray}
p_{ij} = \mbox{Tr}(\rho^s A_{ij})  \neq \mbox{Tr}(\rho^s A_i) \mbox{Tr}(\rho^s A_j) = p_i \, p_j \quad \quad \quad \quad i,j=1,2 \label{qcorr} 
\end{eqnarray}
cannot be given a joint quantum common causal explanation. In (Hofer-Szabó and Vecsernyés, 2012, 2018) we have provided a partition $\{C_k\}$ such that the screening-off conditions (\ref{quantscreen}) hold for all EPR correlations (\ref{qcorr}). That is $\{C_k\}$ is a joint common causal explanation for all the four EPR correlations.\footnote{More than that, we have also shown that in an algebraic quantum field theoretic setting the common causes can even be localized in the common past of the correlating events.}  As also shown in Section 5, these joint common causes need to be \textit{noncommuting} common causes, since commuting common causes would imply the Clauser-Horne inequality (\ref{CH'}) which is violated in the EPR-Bohm scenario. Thus, the violation of the Clauser-Horne inequality (\ref{CH'}) does not exclude a common causal explanation of the EPR-Bohm scenario---as long as noncommuting common causes are tolerated in the explanation.\footnote{The problem with such noncommuting common causes, however, is that it is far from being clear how they should be interpreted.}

To see the thrust of the violation of the Clauser-Horne inequalities, we have to go over to the interpretations of quantum mechanics. As for the operational interpretation, the violation of the Clauser-Horne inequalities, as clarified in Section 3, again does not mean that the numbers (\ref{q1})-(\ref{q3}) cannot be given an operational interpretation. There \textit{is} such an interpretation; otherwise quantum mechanics would not be empirically adequate. The operational interpretation is the following:  
\begin{eqnarray*}
p_i &=& p(A_i \vert a_i) \label{condA}\\
p_j &=& p(B_j \vert b_j) \\
p_{ij} &=& p(A_i \w B_j\vert a_i \w b_j) \label{condAB}
\end{eqnarray*}
where $a_i$ denotes the event that the spin on the left particle is measured in direction $\vec{a}_i$, and $A_i$ denotes the event that the outcome in this measurement is ''up''. (Similarly, for $b_j$ and $B_j$.) That is, the probabilities are conditional probabilities of certain measurement outcomes provided that certain measurements are performed. However, the violation of the Clauser-Horne inequality (\ref{CH'}) does exclude that the conditionally correlating pairs of outcomes  
\begin{eqnarray*}
p_{ij} = p(A_{i}\w A_{j}|a_i \w a_j) \neq p(A_{i}|a_i) \, p(A_{j}|a_j) = p_i \, p_j \quad \quad \quad \quad i,j=1,2 \label{condcorr'} 
\end{eqnarray*}
has a non-conspiratorial joint common causal explanation. Thus, in the operational interpretation, Bell's inequalities filter common causal explanations. 

Finally, let us turn to the ontological interpretation. As is clarified by Pitowsky, the violation of the Clauser-Horne inequality (\ref{CH'}) excludes the numbers (\ref{q1})-(\ref{q3}) to be classical unconditional probability of certain events or properties and their conjunctions. That is, there are no such numbers $A_i$ in a classical probability space such that
\begin{eqnarray*}
p_i &=& p(A_i) \label{A}\\
p_j &=& p(B_j) \\
p_{ij} &=& p(A_i \w B_j) \label{AB}
\end{eqnarray*}
Consequently, there are no correlations 
\begin{eqnarray*}
p_{ij} = p(A_{i}\w A_{j}) \neq p(A_{i}) \, p(A_{j}) = p_i \, p_j \quad \quad \quad \quad i,j=1,2 \label{corr'} 
\end{eqnarray*}
and \textit{a fortiori} no need to look for a common causal explanation. 

As shown in Section 4, a correlation vector has a Kolmogorovian representation if and only if it has a property explanation for any non-signaling conditional Kolmogorovian representation. The probabilities in the EPR-Bohm scenario are non-signaling: 
\begin{eqnarray*}
p_i &=& p(A_i \vert a_i) = p(A_i \vert a_i \w b_j) \\
p_j &=& p(A_i \vert b_j) = p(B_j \vert a_i \w b_j) 
\end{eqnarray*}
Hence, the violation of (\ref{CH'}) again excludes a property explanation for any non-signaling conditional Kolmogorovian representation.

\section{Conclusions}

What does it mean that a set of numbers violates Bell's inequalities? One can answer this question in three different ways depending on whether the numbers are interpreted as classical unconditional probabilities, classical conditional probabilities, or quantum probabilities:
\begin{enumerate}
 \item[(i)] In the first case, the violation of Bell's inequalities excludes these numbers to be interpreted as the classical (unconditional) probabilities of certain events/properties and their conjunctions. The satisfaction of Bell's inequalities does not only guarantee the existence of such events, but also the existence of a set of joint common causes screening off the correlations between these events. 
  \item[(ii)] In the second case, the violation of Bell's inequalities does not exclude that these numbers can be interpreted as the classical conditional probability of certain events (measurement outcomes) conditioned on other events (measurement settings). However, it does exclude a non-conspiratorial joint common causal explanation for the conditional correlations between these events. 
 \item[(iii)] Finally, the violation or satisfaction of Bell's inequalities has no bearing either on whether a set of numbers can be interpreted as quantum probability of certain projections, or whether there can be given a joint common causal explanation for the correlating projections---as long as noncommuting common causes are adopted in the explanation. 
\end{enumerate}

\vspace{0.2in}
\noindent
{\bf Acknowledgements.} This work has been supported by the Hungarian Scientific Research Fund, OTKA K-115593 and a Research Fellowship of the Institute of Advanced Studies Koszeg. I wish to thank Márton Gömöri and Balázs Gyenis for reading and commenting on the manuscript.

\section*{References}
\footnotesize
\begin{list} 
{ }{\setlength{\itemindent}{-15pt}
\setlength{\leftmargin}{15pt}}

\item I. Bengtson and K. \.Zyczkowski, \textit{Geometry of Quantum States: An Introduction to Quantum Entanglement}, Cambridge University Press, Cambridge, 2006.

\item M. Gömöri and T. Placek, ``Small Probability Space Formulation of Bell’s Theorem,'' in G. Hofer-Szabó and L. Wroński (eds.), \textit{Making it Formally Explicit -- Probability, Causality and Indeterminism}, European Studies in the Philosophy of Science Series, Springer Verlag, 109-126 (2017).

\item G. Hofer-Szabó, M. Rédei and L. E. Szabó, \textit{The Principle of the Common Cause}, (Cambridge: Cambridge University Press, 2013).

\item G. Hofer-Szabó and P. Vecsernyés, ''Noncommuting local common causes for correlations violating the Clauser–Horne inequality,'' \textit{J. Math. Phys}, \textbf{53}, 12230 (2012).

\item G. Hofer-Szabó and P. Vecsernyés, ''Bell inequality and common causal explanation in algebraic quantum field theory,'' \textit{Stud. Hist. Phil. Mod. Phys.},  \textbf{44 (4)}, 404–416 (2013).

\item G. Hofer-Szabó and P. Vecsernyés,  \textit{Quantum Theory and Local Causality}, (Dordrecht: Springer Brief, 2018).

\item I. Pitowsky, \textit{Quantum Probability -- Quantum Logic}, (Dordrecht: Springer, 1989).

\item M. Rédei and J. S. Summers, ''Quantum probability theory,'' \textit{Stud. Hist. Phil. Mod. Phys.}, \textbf{38}, 390-417 (2007).

\item L. E. Szabó, ``The Einstein-Podolsky-Rosen argument and the Bell inequalities,'' \textit{Internet Encyclopedia of Philosophy}, URL= http://www.iep.utm.edu/epr/ (2008).
\end{list}

\end{document}